\documentclass[12pt]{article}

\def\Tr{\rm Tr\ }
\def\mt{\tilde{m}}
\def\varphit{\tilde{\varphi}}
\def\sqr#1#2{{\vcenter{\vbox{\hrule height.#2pt
         \hbox{\vrule width.#2pt height#1pt \kern#1pt
            \vrule width.#2pt}
         \hrule height.#2pt}}}}
\def\square{\mathop{\mathchoice\sqr56\sqr56\sqr{3.75}4\sqr34\,}\nolimits}

\begin{document}

\setlength{\baselineskip}{14pt} 
\setlength{\parskip}{1.35ex}
\setlength{\parindent}{0em}

\noindent

\thispagestyle{empty}
{\flushright{\small MIT-CTP-2955\\hep-th/0002226\\}}

\vspace{.3in}
\begin{center}\Large {\bf 
Notes on Fluctuations and Correlation Functions 
in Holographic Renormalization Group Flows}
\end{center}

\vspace{.1in}
\begin{center}
{\large Oliver DeWolfe$^a$ and Daniel Z. Freedman$^{a,b}$}

\vspace{.1in}
{ {\it $^a$Center for Theoretical Physics,\\
Laboratory for Nuclear Science,\\
Department of Physics\\
Massachusetts Institute of Technology\\
Cambridge, Massachusetts 02139, U.S.A. \\ \ \\
$^b$Department of Mathematics \\
Massachusetts Institute of Technology\\
Cambridge, Massachusetts 02139, U.S.A.}}
\vspace{.2in}

E-mail: {\tt odewolfe@mit.edu, dzf@math.mit.edu}
\end{center}
\begin{center}February 2000\end{center}

\vspace{0.1in}
\begin{abstract}
We study the coupled equations describing fluctuations of scalars and
the metric about background solutions of ${\cal N}=8$ gauged
supergravity which are dual to boundary field theories with
renormalization group flow. For the case of a kink solution with a
single varying scalar, we develop a procedure to decouple the
equations, and we solve them in particular examples. However,
difficulties occur in the calculation of correlation functions from
the fluctuations, presumably because the AdS/CFT correspondence has
not yet been properly implemented in the coupled scalar-gravity
sector.  Some new examples of correlators of operators dual to simpler
uncoupled bulk scalars are given and are satisfactory. As byproducts
of our study we make some observations relevant to the stability of
domain walls in the brane-world scenario and to the Hamilton-Jacobi
formulation of holographic RG flows.
\end{abstract}

\newpage
\section{Introduction}

In these notes we study correlation functions in certain
four-dimensional field theories by examining fluctuations about kink
solutions of gauged 5-dimensional ${\cal N}=8$ supergravity
\cite{GRW,PPvN}. In light of the AdS/CFT correspondence
\cite{juan,gkp,witten1}, these gravitational backgrounds have been
interpreted as the duals of ${\cal N}=4$ Super Yang-Mills theory
either perturbed by various relevant operators \cite{gppz1, dz1,
fgpw1, gppz, dz2}, or given expectation values for some scalar fields
\cite{fgpw2}.  Two-, three- and four-point correlation functions
in pure ${\cal N}=4$ SYM are by now well-understood \cite{gkp,
witten1, fmmr, dmmr}, while in kink backgrounds a few two-point
functions have been calculated for operators whose dual scalars are
inert \cite{fgpw2}.  Here we focus on the more complex case of field
theory operators whose dual supergravity scalars are varying radially
in the kink background, although we obtain some new results for inert
scalars also.

Implicit in our use of the AdS/CFT correspondence is the assumption
that 5D maximally supersymmetric gauged supergravity arises as a
consistent truncation of ten-dimensional Type IIB supergravity on
$AdS_5 \times S^5$.  In the case where the dual field theory is on the
Coulomb branch, the ``lifts'' of certain five-dimensional solutions to
ten-dimensional type IIB supergravity are known and correspond to
continuous distributions of D3-branes in the compact extra dimensions
\cite{fgpw2,bs}.  Other ten-dimensional configurations of non-coincident
D-branes have also been studied \cite{kw, cr, rv}. However, we shall
largely take a five-dimensional viewpoint.

Five-dimensional ${\cal N}=8$ gauged supergravity possesses 42 scalar
fields $\varphi^I$, and a potential $V(\varphi)$ which depends on
40 of them.  The scalars fall into various representations of the
$SO(6)$ gauge group: there is a $20'$ representation dual to the
dimension 2 primary operators ${\Tr} X^2$, a $10 \oplus \overline{10}$
representation corresponding to their dimension 3 descendents, and
finally the singlet axion-dilaton, which does not enter into
$V(\varphi)$ and is dual to the ${\cal N}=4$ Lagrangian ${\cal L} =
{\Tr} F^2 + \cdots$, and thus corresponds to the exactly marginal
coupling $\theta + 4\pi i/g^2$.

The part of the supergravity action essential for our concerns is
\begin{eqnarray}
\label{Action}
S = \int d^5x \, \sqrt{g} \left[ - \frac{1}{4} R + \frac{1}{2} \,
g^{\mu \nu} \, \partial_{\mu} \, \varphi^I \, \partial_{\nu} \,
\varphi^I - V(\varphi) \right] \,,
\end{eqnarray}
where the signature of the bulk metric is $(+----)$. In these
conventions the scalar fields are dimensionless.  Tractable kink
solutions typically have 1 or 2 ``active'' scalars $\varphi^I(r)$,
which depend on the radial coordinate $r$, while the remaining
``inert'' scalars $\varphi^\alpha$ are constant (and typically
vanish). The bulk geometry has an $r$-dependent scale factor, and the
kink background takes the form
\begin{eqnarray}
\label{Background}
ds^2 &=& e^{2 A(r)} \, (\eta_{ij} \, dx^i \, dx^j) - dr^2 \,, \\
\varphi^I &=& \varphi^I(r) \,, \nonumber
\end{eqnarray}
which is invariant under 4-dimensional Poincar\'e transformations of
the $x^i, i=0,1,2,3$. The geometries are asymptotically like $AdS_5$,
{\em i.e.}  $A(r) \sim r/L$ as $r \rightarrow \infty$. In most cases
there is a curvature singularity in the interior, {\em i.e.}
$A(r)\rightarrow -\infty$ as $r\rightarrow r_s$. The significance of
the singularity will be briefly discussed below.

We are primarily interested in fluctuations and correlation functions 
of SYM operators
${\mathcal O}_I(x)$ dual to the bulk scalars $\varphi^I(x)$.  Due to
the reduced symmetry in the kink geometries, only two-point
correlators are currently tractable to analysis.  For inert scalars
$\varphi^\alpha$, the procedure is straightforward in principle. One
must obtain solutions of the equations of motion of the fluctuations
$\tilde{\varphi}^\alpha(r,x)$, governed by the action (\ref{Action})
expanded to second order in $\varphit^\alpha$.  Notice that the
effective mass $U(\varphi^I)$ depends on the active scalars:
\begin{eqnarray}
\label{QuadAction}
S_\alpha[\varphi_\alpha] = \frac{1}{2} \int d^5x \sqrt{g} \, \left(
g^{\mu \nu} \partial_{\mu} \varphit^\alpha \partial_{\nu}
\varphit^\alpha - U(\varphi^I(r)) \, \varphit^\alpha{}^2 \right) \,.
\end{eqnarray}
The solution $\varphit^\alpha$ must be chosen to vanish at the
singularity, $r=r_s$, and obey Dirichlet boundary conditions
$\varphit^\alpha(R,x)=\Phi_\alpha(x)$ at some large cutoff radius
$R$. The action (\ref{QuadAction}) is then interpreted as the
generating functional for the field theory.  Evaluated on the
solutions to the equations of motion it reduces to a boundary term,
and the correlation function is obtained from the $R\rightarrow
\infty$ limit,
\begin{eqnarray}
\label{InertCorr}
\tilde{S}_\alpha[\varphi_\alpha] = \frac{1}{2} \int d^4x \sqrt{g} \,
g^{rr} \varphit^\alpha (R,x) \, \partial_r \, \varphit^\alpha (R,x)
\,,
\end{eqnarray}
as prescribed for two-point functions in pure $AdS$ by
\cite{gkp,fmmr}.

Since $U(\varphi^I(r))$ vanishes for the dilaton, the dilaton
two-point function is usually the simplest to compute. Analyticity
properties (in the momentum $p^i$ conjugate to $x^i$) reveal the
spectrum of the boundary theory in the large $N$, large $\lambda=g^2N$
limit. For the various Coulomb branch flows of \cite{fgpw2, bs}, one finds
either a mass gap and continuous spectrum or an infinite discrete
spectrum of poles. Since the supergravity description formally breaks
down at the singularity, it is arguable whether these are actual
features of the field theory dynamics or artifacts of a poor
approximation. We have little to add to these arguments except to
point out that a similar discrete spectrum has also been found for a
distribution of D3-branes with non-singular metric \cite{cr}. Some new
examples of correlators of the dilaton and other inert scalars will be
given below.

For active
scalars the situation is more complex because the fluctuations
$\varphit^I(r,x)$ are coupled by the equations of motion to metric
perturbations $h_{ij}(r,x)$, defined by
\begin{eqnarray}
ds^2 &=& e^{2A(r)} \left( \eta_{ij} + h_{ij}(r,x) \right) dx^i dx^j
- dr^2 \,, \\
\nonumber
\varphi^I &=& \bar\varphi^I(r) + \varphit^I(r,x) \,.
\end{eqnarray}
It is especially interesting that the $\varphit_I$ couple to trace
components of $h_{ij}$, {\em i.e.}
\begin{eqnarray}
h_{ij} = \frac{1}{4} \, h(r,x) \, \eta_{ij} - \frac{\partial}{\partial x^i} \frac{\partial}{\partial x^j} \, H(r,x) \,. 
\end{eqnarray}

This is expected if the flow $\bar{g}_{\mu\nu}(r), \bar{\varphi}^I(r)$
describes a relevant deformation of ${\cal N}=4$ SYM.  In this case
the stress tensor satisfies the operator relation
\begin{eqnarray}
\label{TequalsO}
T^i_i(x) = \sum_I \beta^I \, {\mathcal O}_I(x) \,,
\end{eqnarray}
where ${\mathcal O}_I(x)$ is the relevant operator dual to
$\varphi^I$, and $\beta^I$ is the beta-function for its coupling.
Hence it is natural that the trace of $h_{ij}$, which is dual to
$T^i_i$, should not be independent of the $\varphi^I$.  One might hope
that the present line of investigation interfaces with more general
studies of holographic RG flows \cite{bk,dbvv} and perhaps with ideas
concerning the c-theorem, either holographic \cite{gppz1,fgpw1,agpz} or 
field-theoretic \cite{fl,anselmi,acs}.

On the other hand for a flow describing the Coulomb branch of ${\cal
N}=4$ SYM, one expects that $T_i^i(x)=0$, since the trace vanishes as
an operator if the Lagrangian contains only marginal couplings.  One
then hopes that despite the fact that $\varphit^I$ and $h^i_i$ are
coupled, $h^i_i$ will not excite $T^i_i$.  Since it is well-known that
scalars dual to vevs fall off more rapidly on the boundary than those
dual to operator perturbations, it is reasonable to expect that the excitation
of $h^i_i$ caused by a fluctuation $\varphit^I$ scales away too
quickly to excite an operator in the field theory.  

In the next section we derive the coupled linear fluctuation equations
of $\varphit^I$, $h$, and $H$. Other components of $h_{ij}$ decouple
consistently; the transverse traceless modes are known each to obey
the same wave equation as the dilaton, and the remaining components
can be set to zero by a choice of coordinates.  We discuss a pure
diffeomorphism solution to the equations, the universal solution.  A
more detailed analysis of the coupled fluctuation equations is then
undertaken.  In the case where the background flow involves a single
active scalar $\bar{\varphi}(r)$, we derive an uncoupled third order
differential equation for its fluctuation $\tilde{\varphi}(r,x)$.  The
universal solution allows us to apply reduction of order methods to
study this equation. It is also instructive to convert the fluctuation
equations to equivalent Schr\"odinger form where positivity properties
of supersymmetric quantum mechanics can be applied.  Indeed it turns
out that the role of SUSY QM is ubiquitous.

In the following two sections we solve the fluctuation equations in
detail for two examples of RG flows in which the background is
explicitly known. In section \ref{sym} we treat the flow obtained in
\cite{gppz}, which was interpreted as a relevant perturbation of
${\cal N}=4$ SYM leading to pure ${\cal N}=1$ SYM theory in the
infrared.  In section \ref{coulomb}, we discuss the Coulomb branch
flow of \cite{fgpw2} corresponding to a distribution of D3-branes in a
2-dimensional disc. In both cases we discuss the two-point correlators
of the dilaton\footnote{The dilaton correlator for the Coulomb branch
flow was calculated in \cite{fgpw2}, while for the N=1 flow it was
obtained in the very recent \cite{agpz} which appeared during the
preparation of this manuscript.  } and of a second inert scalar, and
obtain a consistent picture of the spectrum of excitations in the
boundary field theory.  The fluctuation equation for active scalars is
reduced to a hypergeometric equation and solved in both cases, and the
metric component $h(r,x)$ is also determined.

At this point we attempt to determine the two-point correlation
function of the active scalars from (\ref{InertCorr}) by the standard
procedure of identifying the most singular term as $R \rightarrow
\infty$ which is non-analytic in the momentum $p_i$ conjugate to
$x^i$, but encounter difficulties. For the ${\cal N}=1$ flow, the
standard procedure does not have the expected structure of leading
non-analytic term plus more singular polynomials in $p^2$. This
remains true even if an integration constant initially chosen so that
the fluctuation vanishes at the curvature singularity is allowed to
vary. For the Coulomb branch flow, it appears that if the integration
constant is chosen so the active scalar fluctuation is regular, the
metric perturbation $h^i_i$ does not vanish on the boundary as the
argument above for vanishing $T^i_i$ suggests. If the integration
constant is chosen to make $h^i_i$ behave as expected on the boundary,
then we can extract a correlator which is physically reasonable except
for $p^2=0$ poles.  However, regularity of the fluctuations in the
interior then fails.

It is conceivable that the interior curvature singularity is the
source of the difficulties above, but it is also possible that the
standard procedure used to calculate correlators must be modified in
the coupled active scalar/graviton sector. We discuss our attempts to
do this in Sec \ref{2point}. This includes an evaluation of the
on-shell action (\ref{Action}), which reduces to boundary terms linear
and quadratic in fluctuations.  Expected supplementary boundary terms
in the gravitational action are added. An issue of diffeomorphism
invariance is also discussed. None of this leads to a resolution of
the problem, which is left for future work, perhaps by people who can
approach it with fresh energy and ideas.

It is worthwhile to point out two byproducts of our analysis of the
fluctuation equations.  First, the boundary values of the metric and
active scalar fluctuations are not independent, as shown in equation
(\ref{algh}).  It is not clear that this constraint has been
incorporated in a recent Hamilton-Jacobi analysis of holographic RG
flows \cite{dbvv}.  Second, the SUSY QM analysis of the final
uncoupled active scalar equations appears to be applicable to the
stability of domain walls in the brane-world scenario \cite{rs1, rs2,
gw, dfgk}.

\section{The coupled graviton/scalar system}
\label{equations}

We first review the general structure of kink solutions of 5D ${\cal
N}=8$ gauged supergravity which preserve partial supersymmetry. We
then discuss the equations obeyed by linear fluctuations of the kink
backgrounds with particular attention to the equations which couple
fluctuations $\varphit^I(r,x)$ of the active scalars $\varphi^I(r)$.
Fluctuations of inert scalars and transverse traceless metric
fluctuations satisfy uncoupled equations of a simpler structure.

\subsection{Background Flows}

The scalar field manifold of the supergravity theory can be described
as the coset $E_{6(6)}/USp(8)$. The potential $V(\varphi)$ is
invariant under $SO(6) \times SL(2, R) \subset E_{6(6)}$, where $SO(6)
\sim SU(4)$ is the gauge group and the $SL(2,R)$ factor corresponds to
the axion-dilaton. The potential is a very complicated function of 40
scalars, and progress has come from restrictions to smaller sets of
fields which preserve some subgroup $H \subset SO(6)$. Symmetry
properties insure that such restrictions are consistent with the full
dynamics \cite{warnerold}.  The full theory also has a complicated
coset metric $G_{IJ}(\varphi)$ which simplifies to $\delta_{IJ}$ on
the restricted scalar subspaces. This metric was therefore omitted in
(\ref{Action}).

The critical points of $V(\varphi)$ are significant for the problem of
 RG  flows, and  those  extrema  which preserve  at  least an  $SU(2)$
 subgroup of  $SO(6)$ have  been classified \cite{kpw}.  The potential
 takes  negative values  at these  extrema,  so the  solutions of  the
 equations include exact anti-de  Sitter geometries with scalars fixed
 at their  critical values.  In particular there  is a  critical point
 with  full $SO(6)$ symmetry  at the  origin of  field space,  and the
 corresponding $AdS_5$  solution with $SO(4,2)$ isometry  group is the
 well known holographic  dual of the conformal phase  of ${\cal N}=4$,
 d=4  SYM theory. Kink  solutions of  the form  (\ref{Background}) are
 topologically like $AdS_5$ but have  a smaller isometry group. As $r$
 approaches the  boundary at $r=  \infty$, the kink  metric approaches
 that of the maximally  symmetric $AdS_5$ solution, and scalars vanish
 at rates determined  by the scale dimensions of  their operator duals
 in  the SYM  theory.  As $r$  decreases  toward the  interior of  the
 geometry  certain  known  kinks  \cite{fgpw1} flow  toward  a  second
 critical  point  of the  potential  with  more negative  cosmological
 constant. The field theory interpretation is that of ${\cal N}=4$ SYM
 perturbed with a relevant operator such that the perturbed theory has
 a non-trivial IR fixed point. It would be very desirable to study the
 fluctuations about  such flows,  but this cannot  be done  at present
 because   the   explicit   form   of   these   kinks   is   not   yet
 known\footnote{Very recently  the 10D lift of the  fixed point itself
 has  been constructed by  \cite{pilchwarner}.}. All  explicitly known
 flows have  a curvature singularity  at some finite $r=r_s$.  This is
 associated  with a  breakdown of  the supergravity  description.  The
 standard  prescription  is  to  impose the  boundary  condition  that
 fluctuations  vanish at  the singularity  and to  proceed  to extract
 correlation      functions     from      the      on-shell     action
 (\ref{InertCorr}). This procedure leads  to problematic results as we
 will see, but we do not yet have a better alternative to propose.

All explicitly known flow solutions are supersymmetric in the sense
that there are associated Killing spinors in the supergravity
theory. It is known that the potential $V(\varphi)$ can be derived
from a superpotential $W(\varphi)$,
\begin{eqnarray}
\label{Superpot}
V(\varphi) = \frac{g^2}{8} \sum_I \left( \frac{\partial W(\varphi)}{\partial \varphi^I} \, \frac{\partial W(\varphi)}{\partial \varphi^I} \right) - \frac{g^2}{3} W(\varphi)^2 \,.
\end{eqnarray}
on restricted subsets of the field space which include the active
scalars $\varphi^I(r)$. Here $g$ is the $SU(4)$ gauge coupling of the
5D supergravity theory.  It has dimensions of 1/length, and it is
related to the length scale $L$ of the boundary $AdS_5$ space and the
cosmological constant $\Lambda$ by
\begin{eqnarray}
g= 2/L \,, \quad \quad    \Lambda=-12/L^2 = 4V(\varphi=0).
\end{eqnarray}

The main simplification of supersymmetric flows is that any solution
of the first order Killing spinor conditions \cite{fgpw1} 
\begin{eqnarray}
\label{First}
\frac{dA(r)}{dr} = - \frac{g}{3} \, W(\varphi) \,, \quad \quad
 \frac{\partial \varphi^I(r)}{\partial r} = \frac{g}{2} \,
 \frac{\partial W(\varphi)}{\partial \varphi^I} \,,
\end{eqnarray}
automatically gives a Poincar\'e-invariant kink solution of the more
complicated second order field equations of the action (\ref{Action}).

It is worth noting that non-supersymmetric solutions of the form
(\ref{Background}) can also be obtained by solving (\ref{First}), with
a choice of $W(\varphi)$ other than the true superpotential of the
theory.  It was conjectured in \cite{dfgk} that all solutions of the
field equations of the form (\ref{Background}) can be obtained this
way.  Recently this was explained in terms of Hamilton-Jacobi theory
\cite{dbvv}.  However, our examples will be flows where $W(\varphi)$
is in fact the true superpotential.

\subsection{Fluctuation equations}

We now consider fluctuations of the scalars and the metric
around such a background geometry.  We use the freedom to choose
coordinates to fix axial gauge,
\begin{eqnarray}
\label{axial}
h_{\mu 5} = 0 \,,
\end{eqnarray}
where $\mu = 0,1,2,3,5$.  With this choice the metric, including
fluctuations, has the form
\begin{eqnarray}
ds^2 &=& e^{2 A(r)} \, \left( (\eta_{ij} + h_{ij}(x, r)) \, dx^i \, dx^j \right) - dr^2 \,. 
\end{eqnarray}
There is still gauge freedom remaining.  We find two classes of
residual diffeomorphisms $\delta g_{\mu \nu} = \nabla_{\mu}
\epsilon_{\nu} + \nabla_{\nu} \epsilon_{\mu}$ preserving axial gauge
(\ref{axial}). There are four-dimensional diffeomorphisms
\begin{eqnarray}
\label{4ddiffeo}
\epsilon_{i} &=& e^{2A(r)} \, \omega_i(x), \quad \quad \epsilon_5 = 0
\,, \\ \delta h_{ij} &=&  \partial_i \, \omega_j(x) +
\partial_j \, \omega_i(x) \,. \nonumber
\end{eqnarray}
Additionally, there is the transformation \cite{tanaka, gkr}
\begin{eqnarray}
\label{diffeo}
\epsilon_5 &=& \epsilon_5(x) \,, \quad \quad \epsilon_i = -
 (\partial_i \epsilon_5(x) )\, \left(e^{2A(r)} \int^r dr' \, e^{-2A(r')}
 \right) \,, \\ \delta h_{ij} &=& - 2 A'(r) \, \eta_{ij} \, \epsilon_5(x) - 2 \, (\partial_i \partial_j \epsilon_5(x)) \left( \int^r dr' \, e^{-2A(r)} \right) \,. \nonumber
\end{eqnarray}
Here we are viewing the result of an infinitesimal coordinate
transformation on the background as a metric fluctuation.

Both active and inert scalars can fluctuate around their backgrounds:
$\varphi^I(x, r) = \bar\varphi^I(r) + \varphit^I(x,r)$.  For ease of
notation, we shall in the future drop the bar and use $\varphi(r)$ to
denote the classical background.  When we discuss inert scalars, whose
background values are zero, we will drop the tilde on the fluctuation.

Einstein's equations for linear fluctuations were derived in \cite{dfgk}
as equations relating the first order Ricci tensor to its scalar source.
These equations are
\begin{eqnarray}
\label{rij}
``R^{(1)}_{ij}\mbox{''} = e^{2 A} \left(\frac{1}{2} \partial_r^2 + 2
A' \partial_r \right) h_{ij} + {1 \over 2} \, \eta_{ij} \, e^{2A} A'
\, \partial_r (\eta^{kl} h_{kl}) - {1 \over 2} \square h_{ij} - \\
\frac{1}{2} \, \eta^{kl} \left( \partial_i \partial_j h_{kl} -
\partial_i \partial_k h_{jl} - \partial_j \partial_k h_{il} \right) =
-{4 \over 3}\, e^{2A} \, {\partial V(\varphi) \over \partial
\varphi^I} \, \tilde{\varphi^I}\, \eta_{ij} \,, \nonumber
\end{eqnarray}
\begin{eqnarray}
\label{r55}
R^{(1)}_{55} = - {1 \over 2} (\partial_r^2 + 2 A'
\partial_r) \, \eta^{kl} h_{kl} = 4 \,  \varphi^I{}' \tilde{\varphi}^I{}' 
+ {4 \over 3}
  {\partial V(\varphi) \over \partial \varphi^I} 
   \tilde{\varphi^I} \,, 
\end{eqnarray}
\begin{eqnarray}
\label{rj5}
R^{(1)}_{j5} = {1 \over 2}\, \eta^{kl}\, \partial_r\, (\partial_k h_{jl} - \partial_j
h_{kl})  = 2 \, \varphi'^I \, \partial_j \tilde\varphi^I \,.
\end{eqnarray}
The notation $``R^{(1)}_{ij}$'' indicates that a simplification of the
 actual $R^{(1)}_{ij}$ equation has been made in (\ref{rij}) as in
 \cite{dfgk}.  The scalar fluctuation equation is
\begin{eqnarray}
\label{kg}
e^{-2A} \, \square \, \tilde{\varphi}^I - \tilde{\varphi}^I{}'' - 4 A' 
 \tilde{\varphi}^I{}' +  {\partial^2 V(\varphi) \over \partial \varphi^I \partial \varphi^J} 
  \,  \tilde{\varphi}^J = {1 \over 2} \varphi^I{}'
  \eta^{ij} h_{ij}' \,.
\end{eqnarray}
A prime above denotes $\partial_r$.  In general primes on fields
$h_{ij}, \varphit$ and backgrounds $A, \varphi$ will denote a derivative
with respect to the radial coordinate, while primes on $V(\varphi)$
or $W(\varphi)$ refer to $\varphi$-derivatives.

For inert scalars $\varphit^{\alpha}$, the right-hand side of
(\ref{kg}) vanishes, and consequently they do not couple to the
graviton.  Any coupling to the active scalars due to the potential
term must vanish, otherwise the inert scalars could not have been zero
in the background:
\begin{eqnarray}
\frac{\partial^2 V(\varphi)}{\partial \varphi^{\alpha} \partial
\varphi^I} = 0 \,.
\end{eqnarray}
Thus they satisfy decoupled second-order equations, several of which we consider
explicitly in sections \ref{sym} and \ref{coulomb}.

It is clear from the equations above that active scalar fluctuations couple 
to the graviton. To simplify the coupled system 
we consider an arbitrary graviton fluctuation, decomposed in
a complete momentum-space basis.  We use the vectors
\begin{eqnarray}
p^i = (p, 0, 0, 0) \,, \quad \quad \varepsilon^0 = (0,0,0,1) \,, \quad \quad \varepsilon^{\pm} = (0, 1, \pm i, 0) / \sqrt{2} \,,
\end{eqnarray}
and the decomposition
\begin{eqnarray}
h_{ij}(r,p) &=& \varepsilon_i^+ \varepsilon_j^+ \, h^{++}(r,p) +
\varepsilon_i^- \varepsilon_j^- \, h^{--}(r,p) + (\varepsilon_i^+
\varepsilon_j^0 + \varepsilon_i^0 \varepsilon_j^+) \, h^{+0}(r,p) +
\nonumber \\ && (\varepsilon_i^- \varepsilon_j^0 + \varepsilon_i^0
\varepsilon_j^-) \, h^{-0}(r,p) + (\varepsilon_i^+ \varepsilon_j^- +
\varepsilon_i^- \varepsilon_j^+ - 2 \varepsilon_i^0 \varepsilon_j^0)
\, h^{00}(r,p) + \\ && (p_i \varepsilon_j^+ + \varepsilon_i^+ p_j) \,
a^+(r,p) + (p_i \varepsilon_j^- + \varepsilon_i^- p_j) \, a^-(r,p) + (p_i
\varepsilon_j^0 + \varepsilon_i^0 p_j) \, a^0(r,p) + \nonumber \\&&
 \eta_{ij} \, h(r,p) / 4 + p_i \, p_j \, H(r,p) \,.  \nonumber
\end{eqnarray}
In this decomposition, the five $h^{xx}$ are transverse traceless, the
$a^x$ are traceless and longitudinal, and $h$ and $H$ are the trace
components.

The $h^{xx}$ contribute only in (\ref{rij}), and it can be shown that
each $h^{xx}$ satisfies the same uncoupled equation as a free massless
scalar in the background (\ref{Background}), as is well-known
\cite{dfgk, bs2, glueball}. The $h^{xx}$ modes are the expected five physical
components of a graviton in five bulk dimensions, and they have the
same fluctuation spectrum as the inert dilaton. This can be obtained
from an equivalent Schr\"odinger equation with supersymmetric
potential \cite{dfgk}.  These modes do not couple to the active
scalars.

Further, we can examine equation (\ref{rj5}) and see that the $a^x$
must be $r$-independent.  One may then gauge them to zero using the
residual coordinate invariance (\ref{4ddiffeo}).  One is still free to
use the transformation (\ref{diffeo}).

We are left with the components $h$ and $H$, which do couple to active
scalar fluctuations.  Before we present the equations of
this system, consider the action of the remaining gauge freedom
(\ref{diffeo}).  A transformation parameterized by $\epsilon_5(x) =
e^{ipx} \, E(p)$ modifies the fields by
\begin{eqnarray}
\label{Universal}
\delta h = - 8 A'(r) \, E(p) \,, \quad \delta H' = 2 \, e^{-2A(r)} \,
E(p)\,, \quad \delta \varphit = - \varphi'(r) \, E(p)\,.
\end{eqnarray}
One may check that (\ref{Universal}) solves the equations of motion
(\ref{rij}--\ref{kg}).  We will refer to
this pure gauge solution as the {\em universal solution}.  It will
have a vital technical role in our study of the coupled system.

The fields that remain in our reduced system are
\begin{eqnarray}
\label{ansatz}
h_{ij} (r,x) = e^{ipx} \left( \frac{1}{4} \, h(r,p) \, \eta_{ij} + p_i
\, p_j H(r,p) \right) \,, \quad \varphit^I(r, x) = e^{ipx}
\varphit^I(r,p) \,.
\end{eqnarray}
and we now begin the process of simplifying the coupled equations which they
satisfy.

Substituting the ansatz (\ref{ansatz}) into the $R_{j5}$ equation
(\ref{rj5}), $H$ drops out and we are left with
\begin{eqnarray}
\label{rj52}
h'(r) = - \frac{16}{3} \, \varphi^I{}' \, \varphit^I \,,
\end{eqnarray}
where we have canceled a uniform factor of $p_j$.  

We now consider linear combinations of the equations (\ref{rij})
traced with the tensors $\eta^{ij}$ and $p^i \, p^j$.  It turns out
that scalars decouple in the difference of the $\eta^{ij}$ trace and
the $p^i \, p^j$ trace. Dividing by $p^2$ leaves a simple equation
relating $H$ and $h$:
\begin{eqnarray}
\label{Handh}
H'' + 4 A' H' = - \frac{1}{2} \, e^{-2A} \, h \,.
\end{eqnarray}
One can also show that an independent linear combination of the two traces
is trivial given the $R_{j5}$
equation (\ref{rj52}), so the only new relation obtained is
(\ref{Handh}).

Substituting the ansatz (\ref{ansatz}) into the $R_{55}$
equation (\ref{r55}), we obtain
\begin{eqnarray}
\label{r552}
- {1 \over 2} (h'' + 2 A' h' + p^2 H'' + 2 p^2 A' H') = 4 \,
  \varphi^I{}' \, \tilde{\varphi}^I{}' + {4 \over 3} \, \partial_I
  V(\varphi) \, \tilde{\varphi}^I \,.
\end{eqnarray}
Combining (\ref{Handh}) and (\ref{r552}), we can eliminate $H''$ and
obtain an algebraic relation for $H'$.  Further removing derivatives
of $h$ in favor of $\varphit^I$ using (\ref{rj52}) and substituting
$\varphi$-derivatives of $W$ for $\varphi'$ and $V'(\varphi)$, we obtain:
\begin{eqnarray}
\label{algH}
2 A' H' = - \frac{1}{2} \, e^{-2A} \, h + \frac{2g \, \partial_I
W(\varphi)}{3 p^2} \left[ 2 \, \varphit^I{}' - g \, \partial_I
\partial_J W(\varphi) \, \varphit^J \, \right] \,.
\end{eqnarray}

The equations (\ref{rj52}), (\ref{Handh}), (\ref{algH}), together with
the Klein-Gordon equation (\ref{kg}), constitute $n+3$ equations for
the $n+2$ fields $h,H$ and $\varphit^I$.  However, one can show that
the gravitational equations are related by the Bianchi identity: the
expected combination of derivatives of (\ref{rj52}), (\ref{Handh}),
(\ref{algH}) vanishes identically if the Klein-Gordon equation is
satisfied.

It also turns out that we can construct an algebraic equation for $h$ 
which will be of later use.
The Klein-Gordon equation (\ref{kg}) becomes
\begin{eqnarray}
\label{kg2}
- p^2 \, e^{-2A}  \, \tilde{\varphi}^I - \tilde{\varphi}^I{}'' - 4 A' 
 \tilde{\varphi}^I{}' +  {\partial^2 V(\varphi) \over \partial \varphi^I \partial \varphi^J} 
  \,  \tilde{\varphi}^J = {1 \over 2} \varphi^I{}'
  \left( h' + p^2 H' \right)\,.
\end{eqnarray}
The left-hand side is just derivatives of $\varphit^I$.  On the right,
$h'$ can be written in terms of $\varphit^I$ using (\ref{rj52}), while
we can eliminate $H'$ in favor of $\varphit$ and a factor of $h$
using (\ref{algH}).  Thus $h$ is determined solely by $\varphit^I$ and
derivatives.  To avoid tedious notation, we write only the case of a
single active scalar, although the general case is straightforward:
\begin{eqnarray}
\label{algh}
 3 p^2 e^{-2A} \, h = \frac{-16 W} {W'}
\varphit'' + g \left( \frac{64 W^2}{3 W'} + 8
W' \right) \varphit' + \\ 4 g^2 \left( \frac{
W W''^2}{W'} +  W W'''(\varphit) - \frac{8 W^2
W''}{3W'} -  W'
W'' - \frac{4 p^2 W}{g^2 W'} e^{-2A}
\right) \varphit \,.
\nonumber
\end{eqnarray}
This equation shows that the boundary values of $h(r,p)$ and $\varphit(r,p)$
are not independent, a fact which may have implications for the formulation
of Hamilton-Jacobi dynamics proposed in \cite{dbvv}.

Unfortunately our equations are still coupled. The next step is to
derive a third order equation involving only the $\varphit^I$. In the
case of flows with only a single active scalar, this becomes an
uncoupled equation which is the key to the present exploration and can be
solved for some specific flows. 

The Klein-Gordon equation (\ref{kg}) has the conventional form of a
scalar field in the curved background, with an additional source term
involving the graviton fluctuation.  We can express the source in
terms of scalars by noting that the $R_{55}$ equation (\ref{r55}) can
be written
\begin{eqnarray}
- \frac{1}{2} e^{-2A} \partial_r (e^{2A} \eta^{ij} h_{ij}') = 4 \varphi^I{}'
 \tilde{\varphi}^I{}' + {4 \over 3} {\partial V(\varphi) \over \partial \varphi^I}
 \, \tilde{\varphi}^I \,,
\end{eqnarray}
which can be integrated to obtain
\begin{eqnarray}
\eta^{ij} h_{ij}' = - 8 \, e^{-2A} \int^r \, dr' \, e^{2A(r')} \left(
\varphi^I{}' \, \varphit^I{}' + \frac{1}{3} \, \partial_I  V(\varphi) \, \varphit^I \right) \,.
\end{eqnarray}
Substituting into (\ref{kg}) and taking an $r$-derivative,
we find the third-order equation 
\begin{eqnarray}
\label{thirdmany}
\partial_r \, \left\{ \frac{\varphi^I{}'}{|\varphi^K{}'|^2} \, e^{2A} \,
\left( \left( \partial_r^2 + 4A' \partial_r- e^{-2A} \square \right) \,
\delta_{IJ} - \partial_I \partial_J V(\varphi) \right) \, \varphit^J \right\} = \\
4 \, e^{2A} \left( \varphi^J{}' \varphit^J{}' + \frac{1}{3} \partial_J
V(\varphi) \, \varphit^J \right) \,. \nonumber
\end{eqnarray}

Although a simplification, (\ref{thirdmany}) still couples all active
scalar fluctuations.  Since several known flows involve only one
active scalar, we
specialize to the case of a single active scalar $\varphi$. 
Then (\ref{thirdmany}) reduces to an uncoupled equation,
\begin{eqnarray}
\label{third}
\partial_r \, \left\{ \frac{1}{\varphi'} \, e^{2A} \, \left( \partial_r^2 +
4A' \partial_r - V''(\varphi) - e^{-2A} \square \right) \, \varphit \right\}
= 4 \, e^{2A} \left( \varphi' \varphit' + \frac{1}{3} V'(\varphi) \, \varphit
\right) \,.
\end{eqnarray}

In the case of a single active scalar, it is also possible to combine
(\ref{rj52}, \ref{Handh}, \ref{algH}) to obtain an uncoupled third order
equation for $h$, but we have chosen to work with (\ref{third}) in
applications to specific flows.

One may easily verify that the universal solution, $\varphit \propto
W'(\varphi) \, e^{ipx}$, satisfies (\ref{third}) by using
(\ref{Superpot}) and (\ref{First}) to relate $V(\varphi)$, $\varphi'(r)$ and
$A'(r)$ to the superpotential $W(\varphi)$ and its derivatives. 
This allows us to use the method of
reduction of order to obtain a second-order equation for the remaining
solutions.  Writing
\begin{eqnarray}
\label{tilde}
\varphit = W'(\varphi) \, e^{ipx} \, \int^r dr' \, R(r',p) \,,
\end{eqnarray}
for some unknown $R(r,p)$, and using the properties of the flow
(\ref{Superpot}) and (\ref{First}), we find
\begin{eqnarray}
\label{second}
R''(r,p) + g \, (W'' - 2W) \, R'(r,p) &+& \\
g^2  \left( -\frac{2}{3} W W'' + \frac{8}{9}
W^2 + \frac{1}{2} W' W''' - W'^2 \right) \, R(r,p) &+& p^2 \, e^{-2A} \,
R(r,p) = 0 \,. \nonumber
\end{eqnarray}
We remind the reader that primes on $R(r)$ are $r$-derivatives while
those on $W$ are with respect to $\varphi$. To determine the boundary
scaling rates of the three independent solutions of (\ref{third}), we
may approximate $W(\varphi) = -3/2 + \rho \, \varphi^2/2$ where
$\rho$ is related to the operator scale dimension $\Delta$ by
$\rho=\Delta-4$ for operator deformations and $\rho=-\Delta$ for
vacuum expectation values. Standard Frobenius analysis then gives the
exponential rates
\begin{eqnarray}
\varphit(r) \sim \exp(\rho r/L), \quad \exp((\rho-2)r/L), \quad
\exp((-\rho-4)r/L) \,,
\end{eqnarray}
for the universal solution, and the two independent solutions of
(\ref{second}), respectively.

By solving (\ref{second}), we can obtain the solution to $\varphit$
from (\ref{tilde}).  The integration constant is just the freedom to
add a multiple of the universal solution to our result for $\varphit$.
This will be important later on for assuring regularity of the
solutions at the singularity.

\subsection{Supersymmetric Quantum Mechanics}

We now discuss the transformation of the fluctuation equations into
Schr\"odinger form, which gives considerable intuition \cite{fgpw2}
into the nature of the fluctuation spectrum.  For the dilaton and for
the transverse traceless metric fluctuations \cite{dfgk,bakas} the
Schr\"odinger potential is that of a supersymmetric quantum mechanics,
and this gives a simple proof that the spectrum of normalizable
fluctuations contains no tachyons.  We will see that SUSY-QM provides
a useful framework in which to consider all fluctuations.

The Schr\"odinger form obtains after a change of radial coordinate $r$
to the horospherical coordinate $z$, defined by the line element
\begin{eqnarray}
ds^2 = e^{2A(z)} \, \left[ (\eta_{ij} + h_{ij}(x,z)) \, dx^i dx^j - dz^2 \right] \,,
\end{eqnarray}
and thus the Jacobian must be
\begin{eqnarray}
\label{horo}
\frac{dz}{dr} = \pm \, e^{-A} \,.
\end{eqnarray}
We choose the $-$ sign for the present application, so that the
boundary region $r \rightarrow \infty$ is mapped to $z=0$ and the deep
interior to $z \rightarrow \infty$.

Consider first the fluctuation equation for an inert scalar
$\varphit(r,p)$ obtained from (\ref{QuadAction}),
\begin{eqnarray}
\label{inerteq}
- \left( \square + U(\varphi) \right)  \varphit(r,p) = \varphit''(r,p) 
+ 4 A'(r) \, \varphit'(r,p) - U(\varphi(r)) \, \varphit(r,p) + e^{-2A}
 \, p^2 \, \varphit(r,p) = 0 \,.  
\end{eqnarray}
Performing the field transformation $\varphit(r,p) = \exp(-3A/2) \,
\psi(z,p)$ as well as the change of coordinate $r \rightarrow z$, one
finds an equation in Schr\"odinger form
\begin{eqnarray}
\label{scheq}
- \psi''(z,p) + {\cal V}(z) \, \psi(z,p) = p^2 \psi(z,p) \,,
\end{eqnarray}
with potential
\begin{eqnarray}
\label{schpot}
{\cal V}(z) = \left( \frac{3}{2} A'(z) \right)^2 + \frac{3}{2} A''(z) +
e^{2A(z)} \, U(\varphi(z)) \,,
\end{eqnarray}
in which $A'(z) = dA/dz = \pm \exp(A) \, A'(r)$.  (Note that either
sign choice in (\ref{horo}) leads to the same form for (\ref{scheq})
and (\ref{schpot}).)  The first two terms in (\ref{schpot}) are in the
form of a supersymmetric quantum mechanics potential ${\cal V}(z) =
{\cal U}(z)^2 + {\cal U}'(z)$ derived from a prepotential ${\cal U}(z)
= (3/2) \, A'(z)$, while $U(\varphi(z))$ is determined by the coupling
of inert and active scalars in the supergravity potential.  If
$U(\varphi(z))$ is positive (or absent as it is for the dilaton
fluctuation), then one has an immediate argument that normalizable
fluctuations occur only for $p^2 \geq 0$.  In general, the behavior
of the potential (\ref{schpot}) near the limits $z \rightarrow 0$ and
$z \rightarrow \infty$ is usually enough information to ascertain
whether the fluctuation spectrum is discrete or continuous, with or
without gap \cite{fgpw2}.

In our examples in sections \ref{sym} and
\ref{coulomb}, the two inert fluctuations with nonzero $U(\varphi(z))$
turn out to have $U(\varphi(z)) < 0$, so that positivity of $p^2$ is
not obvious with the potential in the form ($\ref{schpot}$).  However
in both cases we are able to rewrite the potentials in an exact SUSY
QM form with a modified ${\cal U}$.  The two examples work
differently, and we leave the question of the existence of an exact
SUSY QM potential for general coupled inert scalars for the future.

The norm is a potentially delicate issue.  The Hamiltonian in
(\ref{scheq}) is self-adjoint with respect to the ``Schr\"odinger
norm'' $\int dz \, \psi(z)^2$, but not with respect to the transformed
covariant norm $\int dz \, \exp(2A) \, \psi(z)^2$ which is correct in
the present setting.  (This results because a factor of $\exp(-2A)$
was dropped in passing from (\ref{inerteq}) to (\ref{scheq}).)
Because of this delicacy, solutions which have infinite Schr\"odinger
norm but finite covariant norm would not necessarily have $p^2>0$. 

The treatment of active scalar fluctuations is more complex.
The goal is to transform the second order equation (\ref{second}) to
Schr\"odinger form.  The change from $r$ to $z$ produces 
\begin{eqnarray}
\label{schactive}
- R''(z) + g e^A \left( W'' - \frac{5}{3} W \right)  R'(z) +
  g^2 e^{2A}  \left (W'^2 + \frac{2}{3}W W'' - \frac{8}{9} W^2 -
  \frac{1}{2} W' W''' \right) R = p^2 R \,,
\end{eqnarray}
in which $R'$ and $R''$ are derivatives with respect to $z$, but $W'$,
$W''$ etc.\ indicate $\varphi$-derivatives, as has been our practice.
The further transformation
$R(z) = \exp(S(z)) \, \psi(z)$, with
\begin{eqnarray}
\frac{dS}{dz} = \frac{1}{2} \, g \, e^{A(z)}  \left( W''(\varphi(z)) -
\frac{5}{3} W(\varphi(z)) \right) \,,
\end{eqnarray}
produces the Schr\"odinger form (\ref{scheq}) with the (ugly) potential
\begin{eqnarray}
\label{activepot}
{\cal V}(z) = \frac{1}{4} \, g^2 e^{2A(z)} \, \left[ \frac{1}{3} W^2 +
\frac{7}{3} W'^2 + (W'')^2 - \frac{4}{3} W W'' - W' W''' \right] \,.
\end{eqnarray}
Unpromising as it seems, one may try to express (\ref{activepot}) in
SUSY-QM form perhaps with a simple remainder.  To do this we use the
ansatz 
\begin{eqnarray}
{\cal U} = e^A \, (aW +bW' +cW'') \,,
\end{eqnarray}
where $a,b,c$ are free parameters. We find that an exact SUSY form
cannot be achieved, but analogously to ($\ref{schpot})$ one may write
\begin{eqnarray}
\label{activeplusremain}
{\cal V}(z) = {\cal U}(z)^2 + {\cal U}'(z) + \frac{1}{3}\, g^2 e^{2A(z)}
\, (W')^2 \,,
\end{eqnarray}
with prepotential
\begin{eqnarray}
\label{activepre}
{\cal U}(z) = \frac{1}{2} \, g \, e^{A(z)} \, (W'' - W) \,.
\end{eqnarray}
The discussion below
(\ref{schpot}) applies here as well, but here it is manifest that the
additional term beyond the SUSY-QM structure $\Delta {\cal V} = (1/3)
\, e^{2A} \, (W')^2$ is positive-definite.  This indicates that there
are no tachyons in the normalized fluctuation spectrum of the active
scalar.  In particular examples one must be careful to use the correct
norm which is the transformation of $\int dr \, e^{4A(r)}\, \varphit^2$ with
$\varphit(r,p)$ given by (\ref{tilde}).

Two further comments can be made.  Although in general ${\cal V}(z)$
only can be cast in the form (\ref{activeplusremain}), in specific
examples it may well have an exact SUSY form with a prepotential
different from (\ref{activepre}).  An example of this appears in
section \ref{coulombactive}.  Additionally, it is also clear that the
analysis above has applications to the stability of domain
walls in the brane world scenario \cite{rs1,rs2} which we plan to
pursue.

\section{The ${\cal N} = 1$ Super Yang-Mills flow}
\label{sym}

Several kink solutions involving a single real flowing scalar have
been considered in the literature.  There are only two inequivalent
single real scalars transforming in the ${\bf 10} \oplus {\bf
\overline{10}}$ of $SU(4)$, as there are two Weyl orbits in the ${\bf
10}$.  Representative scalars in the longer and shorter orbits,
respectively, have been called $\sigma$ and $m$, and we will use this
convention.  There are three inequivalent choices in the ${\bf 20'}$, one of
which we will discuss in the next section.

Interesting renormalization group flows have been considered in
\cite{gppz} in which the scalar manifold of bulk ${\cal N}=8$ gauged
supergravity was restricted to a subspace of two particular
inequivalent scalars in the ${\bf 10} \oplus {\bf \overline{10}}$ of
$SU(4)$.  A kink solution with only $m(r)$ active was obtained and
interpreted as the holographic dual of a perturbation of ${\cal N}=4$
SYM theory by a dimension 3 operator which leads to a field theoretic
RG flow to pure ${\cal N}=1$ SYM theory at long distances. The
10-dimensional lift of this solution is not known, and the
5-dimensional geometry has an interior curvature singularity. The
significance of this is not clear, but a scenario for its stringy
resolution has been outlined \cite{sp}.

We will use the kink background of \cite{gppz} as a testing ground for
computations of fluctuations and correlation functions as discussed
above.  For this purpose we review the background solution in the next
subsection. We then go on to study fluctuations and correlators of two
inert scalars, and finally focus on perturbations $\mt(r,p)$ of the
active scalar.

\subsection{The background flow}

In the consistent sub-sector of ${\cal N}=8$ supergravity considered in
\cite{gppz}, the two scalar fields $m$ and $\sigma$ have canonical
kinetic terms and superpotential and potential given by:
\begin{eqnarray}
W(m, \sigma) &=& - \frac{3}{4} \left[ \cosh \left( \frac{2m}{\sqrt{3}} \right) + \cosh \left(2 \sigma \right) \right] \,, \\
V(m, \sigma) &=& - \frac{3g^2}{8} \left[  \frac{1}{4} \cosh^2 \left( \frac{2m}{\sqrt{3}} \right) +  \cosh\left(\frac{2
m}{\sqrt{3}}\right) \cosh(2 \sigma) - \frac{1}{4} \cosh^2 (2 \sigma) + 1 
\right] \,. \label{msigmapot}
\end{eqnarray}
Since the $\sigma$ field appears quadratically in $V(m,\sigma)$ there is a
supersymmetric flow in which it vanishes. This flow is a solution of
(\ref{First}) with superpotential
\begin{eqnarray}
W(m) = - \frac{3}{4} \left[1 + \cosh \left(\frac{2m}{\sqrt{3}} \right)
\right] \,.
\end{eqnarray}

In the following, we set $g = 2/L$, and $W$ always refers to $W(m)$.  The kink
solution is
\begin{eqnarray}
m(r) = \frac{\sqrt{3}}{2} \log \frac{1+ e^{-r/L}}{1-e^{-r/L}} \,, \\
A(r) = \frac{1}{2} \left( \frac{r}{L} + \log 2 \sinh \frac{r}{L} \right) \,.
\end{eqnarray}
There is a singularity at finite proper distance, which by choice of
an additive integration constant we have located at $r_s=0$.

The horospheric coordinate $z$ is obtained from
$dr/dz = - e^A$, which can be solved to give
\begin{eqnarray}
e^{2A(z)} = \cot^2 \left( \frac{z}{L} \right) = e^{\frac{2r}{L}} -1 \,.
\end{eqnarray}
Neither $r$ nor $z$ proves useful for solving the various equations.
Instead we are able to make progress by using a variable $u$ in which
the boundary is mapped to $u_b = 1$ and the singularity to $u_s = 0$.
This is achieved by
\begin{eqnarray}
\label{udef}
u \equiv \cos^2 \left( \frac{z}{L} \right) \,.
\end{eqnarray}
The equations for fluctuations become hypergeometric in $u$ as we will see,
and the quantities which enter these equations can be expressed as:
\begin{eqnarray}
\label{symreference}
W(u) &=& - \frac{3}{2u} \,, \quad \quad \;\;
W'(u) = - \sqrt{3} \; \frac{\sqrt{1-u}}{u} \,, \nonumber \\
W''(u) &=&  \frac{u-2}{u} \,, \quad \;
W'''(u) = - \frac{4}{\sqrt{3}} \, \frac{\sqrt{1-u}}{u} \,, 
\\
e^{2A(u)} &=& \frac{u}{1-u} \,, \quad \quad 
\frac{du}{dr} = \frac{2}{L} \, (1-u) \,. \nonumber
\end{eqnarray}

\subsection{Correlators of inert scalars}

Let us begin our study of fluctuations in this geometry by calculating the
two-point function for the dilaton $\phi$.  The dilaton couples to
${\cal O}_\phi ={\Tr} F^2 + \cdots$ and hence provides information 
about the glueball
spectrum.  The Klein-Gordon equation for this field is 
\begin{eqnarray}
- \frac{1}{\sqrt{g}} \, \partial_{\mu} \, (\sqrt{g} \, g^{\mu \nu}
 \partial_{\nu} ) \, \phi = \left(\frac{\partial}{\partial r} +4A'(r)
 \right) \frac{ \partial \phi}{\partial r} +p^2 e^{-2A} \phi =0.
\end{eqnarray}

In terms of $u$ this becomes
\begin{eqnarray}
\phi''(u) + \frac{1}{1-u} \, \left( \frac{2}{u} -1 \right) \, \phi'(u)
+ \frac{p^2 L^2}{4} \, \frac{1}{u (1-u)} \, \phi(u) = 0 \,.
\end{eqnarray}
This equation is hypergeometric; the solution regular in the
deep interior ($u = 0$) is
\begin{eqnarray}
\phi(u) = (u-1)^2 \, F \left( 2- \frac{pL}{2},2+ \frac{pL}{2};2;u \right) \,.
\end{eqnarray}
This is a simpler, but presumably equivalent, form of the solution very 
recently
obtained in \cite{agpz}, where a different independent variable was used.

We can calculate the two-point function in the established way
\cite{gkp,fmmr}. Specifically, the $p$-space correlator is obtained by
imposing a cutoff at $z=\epsilon$ in the horospheric coordinate. One
then finds
\begin{eqnarray}
\label{2ptfunc}
\langle{\cal O}(p) \, {\cal O} (-p)\rangle = - \lim_{\epsilon \rightarrow 0} \, \left( \frac{1}{\epsilon^{2(\Delta-4)}} \right) \left[
\frac{1}{z^3} \frac{d}{dz} \ln(\phi(z,p)) \right]_{z= \epsilon},
\end{eqnarray}

where an energy-momentum conserving $\delta$-function has been dropped,
and $\Delta$ is the UV dimension of the operator ${\cal O}(x)$. The only
term which is kept in the limit is the most singular term in
$\epsilon$ which is non-analytic in $p^2$. Other more singular terms
are multiplied by polynomials in $p^2$, and are dropped because they
correspond to counter terms. This procedure is equivalent to
calculating the second variation of the on-shell action
(\ref{InertCorr}). Field theory considerations imply that correlators
behave as $(-p)^{2\nu} \ln(-p^2)$ for large spacelike momentum, where
$\nu = \Delta - 2$.
Our calculations lead directly to correlation functions
which are normalized so that the leading term has the same coefficient as
in the $AdS_5$ geometry. The analysis of the Appendix of \cite{fmmr}, redone
for Bessel functions with integer $\nu > 2$, gives 
\begin{eqnarray}
\label{pureadscorr}
\langle{\cal O}(p) \, {\cal O} (-p)\rangle = - \left[ \frac{2\nu}{\Gamma(\nu)\Gamma(\nu+1)4^\nu} \right] (-p)^{2\nu} \ln(-p^2)        
\end{eqnarray}
For the dilaton this procedure gives (with $s=p^2$)
\begin{eqnarray}
\langle{\cal O}_{\phi}(p) \, {\cal O}_{\phi}(-p)\rangle = - \frac{1}{8} \,
s \left( s-\frac{4}{L^2} \right) \left[\psi \left( 2 + \frac{L\sqrt{s}}{2} \right) +
\psi \left( 2 - \frac{L \sqrt{s}}{2} \right) \right] \,,
\end{eqnarray}
where $\psi(z) \equiv d\ln \Gamma(z)/dz$. This result agrees with \cite{agpz}.
The
correlator has a discrete spectrum of poles at $s=4(n+2)^2/L^2 $ as is
consistent with the expected glueball spectrum in a confining
theory.

We can also consider the two-point function of the inert
scalar $\sigma$.  One must now take into account the coupling $U(m)$
to the active scalar that results from the potential (\ref{msigmapot}).
The wave equation is (\ref{inerteq}) with
\begin{eqnarray}
U(m) \equiv \left. \frac{\partial^2 V(m,\sigma)}{\partial \sigma^2}
\right|_{\sigma = 0} = - \frac{3 g^2}{4} \left[ 2 \cosh
\left(\frac{2m}{\sqrt{3}} \right) -1 \right] \,,
\end{eqnarray}
which leads to the hypergeometric wave equation
\begin{eqnarray}
\sigma''(u) + \frac{1}{1-u} \, \left( \frac{2}{u} -1 \right) \, \sigma'(u)
+ \frac{1}{(1-u)^2} \left( \frac{p^2 L^2}{4} \, \frac{1-u}{u}
+ \frac{3}{4} \frac{4-3u}{u} \right)
 \, \sigma(u) = 0 \,.
\end{eqnarray}
The solution regular at the horizon is
\begin{eqnarray}
\sigma(u) = (1-u)^{3/2} F \left( \frac{3}{2} + \frac{1}{2} \sqrt{9 +
L^2 p^2}, \frac{3}{2} - \frac{1}{2} \sqrt{9 + L^2 p^2};2;u \right) \,.
\end{eqnarray}
The corresponding two-point function is 
\begin{eqnarray}
\label{sigmacorr}
\langle{\cal O}_{\sigma}(p) \, {\cal O}_{\sigma}(-p)\rangle =
\frac{s+8/L^2}{2} \, \left[\psi \left( \frac{3}{2} + \frac{1}{2}
\sqrt{9 + L^2 p^2} \right) + \psi \left(\frac{3}{2} - \frac{1}{2}
\sqrt{9 + L^2 p^2} \right) \right] \,.
\end{eqnarray}
There is a discrete spectrum of poles at $s=4n(n+3)/L^2, n=0,1,2...$. We
note the presence a massless state, which we shall not attempt to reconcile
with the interpretation \cite{gppz}that the flow describes a confining
field theory.

These results may be compared with expectations based on the form of
the Schr\"odinger potentials (\ref{schpot}) discussed in section
\ref{equations}.  For both fields, the one discovers that ${\cal V}(z)
\rightarrow + \infty$ in both the boundary and deep interior limits,
which is indicative of the discrete spectrum found above.  In section
\ref{equations} we showed that the Schr\"odinger potential of an inert
dilaton always possesses a SUSY QM form ${\cal V}(z) = {\cal U}(x)^2 +
{\cal U}'(z)$, where
\begin{eqnarray}
\label{universaldil}
{\cal U}_{\phi}(z) = \frac{1}{2} g \, e^A \, W \,,
\end{eqnarray}
independent of the flow.  Thus we can argue that the spectrum is
positive, and we are in the situation of broken supersymmetry since
the candidate 0-mode (which is just $\phi = const$) is not
normalizable.  The $\sigma$ field has an additional term proportional
to $U(m(z))$, which turns out to be negative; the $m$-flow sits atop a
ridge in the $\sigma$-direction of field space.  Thus SUSY quantum
mechanics does not manifestly apply with the potential written as in
(\ref{schpot}).  However, one may show that ${\cal V}_{\sigma}(z)$ can
be rewritten in an exact SUSY QM form with prepotential
\begin{eqnarray}
\label{symsigmasusypot}
{\cal U}_{\sigma}(z) = \frac{1}{2} g \, e^A \left( W + 3 \right) \,,
\end{eqnarray}
and in this case the zero-mode can be shown to be normalizable, consistent
with the spectrum of the correlator (\ref{sigmacorr}).

\subsection{The graviton/active scalar system}

It is straightforward to show that (\ref{second}) becomes
\begin{eqnarray}
\label{symR}
R''(u) + \frac{1}{u(1-u)} R'(u) + \frac{1}{u(1-u)} \left( \frac{2u-1}{u(1-u)}+
p^2L^2/4 \right) R(u) = 0 \,,
\end{eqnarray}
which has the solution 
\begin{eqnarray}
\label{symRsoln}
R(u) = u (1-u) F\left( \frac{3}{2} + \frac{1}{2}\, \zeta, \frac{3}{2} - \frac{1}{2}\, \zeta; 3; u \right) \,,
\end{eqnarray}
where $\zeta = \sqrt{1 + p^2 L^2}$.  We reconstruct $\mt$ following
(\ref{tilde}).  Using the relations (\ref{symreference}), we integrate
to find
\begin{eqnarray}
\label{symM}
\mt = \frac{\sqrt{1-u}}{u} \left[ 4 \, F_1 \, (p;u) - f_0  +
p^2 \,L^2 \, u\, F_2\, (p;u)  \right] \,,
\end{eqnarray}
where $F_n \, (p ;u)$ is the hypergeometric function
\begin{eqnarray}
F_n \, (p ; u) \equiv F \left( n - \frac{3}{2} + \frac{1}{2}\, \zeta,  n - \frac{3}{2} - \frac{1}{2}\, \zeta; n;u \right)  \,.     
\end{eqnarray}
Here $f_0$ is an integration constant, corresponding to the addition
of a multiple of the universal solution.  The choice $f_0 = 4$ ensures
regularity at the singularity $u=0$, but we will keep $f_0$ as a
parameter in our further discussion.  The second solution to
(\ref{symR}) is singular at $u=0$ and cannot be made regular.  Near
the boundary $u=1$, we observe $\mt \sim \sqrt{1-u} \sim z$ which is
the proper scaling for an operator with dimension $\Delta = 3$.

Let us check to see whether the form found for the fluctuation agrees
with the intuition based on Schr\"odinger form (\ref{schactive}) of
(\ref{symR}), with potential ${\cal V}(z)$ given by (\ref{activepot}).
Near $z=0$, the leading term of the potential is ${\cal V}(z) =
-1/4z^2$, which is the limiting strength of an allowed attractive
$1/z^2$ potential.  Approaching the singularity at $z \rightarrow
\pi/2$, there is a repulsive $1/(\pi/2 - z)^2$ behavior.  Thus one
expects a positive discrete spectrum, and this is reflected in the
values of $\zeta$ which give a terminating hypergeometric series in
(\ref{symRsoln}), namely $\zeta = 2n +3$ or $s = p^2 = 4 (n+1)(n+2)$ for
$n = 0,1,2\ldots$.  Examining (\ref{symM}) at these values of $p^2$,
we find the behavior near the boundary
\begin{eqnarray}
\left. \mt(u,p) \right|_{\zeta = 2n+3} \approx - \sqrt{1-u} \left[ f_0 + c_n (1-u) + {\cal O}(1-u)^2 \right] \,,
\end{eqnarray}
where $c_n$ is a constant.  These functions are normalizable (in the
covariant norm $\int dr \, e^{4A} \, (\mt)^2$) only if $f_0 = 0$.

We now attempt to obtain a correlation function by applying the previously used standard procedure.  Hypergeometric analytic continuation formulae \cite{bateman} give the boundary behavior
\begin{eqnarray}
\label{Mboundary}
\mt(u,p) &\approx& \frac{\sqrt{1-u}}{u} \, \frac{1}{\Gamma
\left(\frac{3+ \zeta}{2} \right) \Gamma \left(\frac{3- \zeta}{2} \right)}
\left[ 4+p^2+\frac{f_0 \pi p^2}{4 \cos(\pi \zeta/2)} \right. \\ 
&+& \left. \frac{p^4(1-u)}{4} \left(h_0''- \ln(1-u) \right) + {\cal
O}(1-u)^2 \right] \,, \nonumber
\end{eqnarray}
where $h_0'' = \psi(1) + \psi(2) - \psi((3+\zeta)/2)- \psi((3-\zeta)/2)$.
Applying (\ref{2ptfunc}) we see that the ``would-be correlator''
should be the most singular term (in $(1-u)$) that is non-analytic in
$p^2$ in
\begin{eqnarray}
\label{badcorr} ``\langle{\cal O}(p) {\cal O}(-p)\rangle\mbox{''} = \frac{s^2}{4s + 16/L^2 +
\frac{f_0 \pi s}{\cos(\pi \zeta/2)}} \left[ \psi \left(\frac{3}{2} +
\frac{1}{2} \zeta \right) + \psi \left(\frac{3}{2} - \frac{1}{2} \zeta
\right) - \ln\left((1-u)e^{2 \gamma} \right) \right] .
\end{eqnarray}
At this point we hit the first major barrier of our program; we see no
way to extract the leading non-analytic term.  For example the
coefficient of $\ln(1-u)$ should be a polynomial in $s$ in order to be
interpreted as a contact term.  However here we have (for general
$f_0$) an entire function of $s$ with essential singularity at
infinity, while at $f_0 = 0$ there is a pole at the space-like value
$s = - 4/L^2$.  One might nevertheless try to identify the correlator
by simply dropping the log term.  However for general $f_0$ the
expected poles of the discrete spectrum, at $\zeta = 2s+3$, cancel
between the numerator and denominator, leaving a function with no
singularities in the finite $s$-plane.  For $f_0 = 0$ one regains the
expected poles, but there are also unphysical tachyon poles.

We are rather sure that our treatment of the fluctuation equations
leading to (\ref{tilde}), (\ref{second}) is correct.  The solution
$\mt(u,p)$, especially at $f_0=4$, has the properties expected in the
$AdS$/CFT correspondence, both at the boundary and the interior
singularity.  It is reassuring that (\ref{second}) admits solutions
with unphysical behavior in both limits, yet the actual solution
$\mt(r,p)$ behaves correctly at both ends. It thus appears that it is
the procedure to obtain the correlation function from the fluctuation
which must be modified.

We can additionally calculate the associated graviton fluctuations using
equations (\ref{algH}) and (\ref{algh}).  We find
\begin{eqnarray}
\label{symh}
h &=& \frac{1}{3 \sqrt{3} u} \left( 24f_0 - 96 F_1 - 24 L^2 p^2 u
F_2 +24 L^2 p^2 u (1-u) F_3 \,  \right. \\ &+& \left.  L^2 p^2 u^2 (1-u)
(8 - L^2 p^2) F_4 \right) \,, \nonumber \\ \label{symH} H' &=& -
\frac{L(1-u)}{12 \sqrt{3} u} \left( 24f_0 - 96 F_1 - 24 L^2 p^2 u F_2
+12 L^2 p^2 u (1-u) F_3 \,  \right. \\ &+& \left.  L^2 p^2 u^2 (1-u) (8 -
L^2 p^2) F_4 \right) \,. \nonumber
\end{eqnarray}
Only $H'(r)$ is determined by the equations.  Note that if $f_0 = 4$,
$\mt$, $h$ and $H'$ are all regular at the singularity $u = 0$.

In principle, one would like to use (\ref{symh}) and (\ref{symH}) to
calculate correlation functions of the trace of the energy-momentum
tensor, such as $\langle T^{i}_i(p) \,T^j_j(-p)\rangle$, which is related to
renormalization group flows and proposals for c-theorems
\cite{anselmi}.  The fact that the solutions for $\mt$, $h$ and $H$
are all determined by a single set of boundary data is consistent with
the field theory reality that the operator $T^i_i$ is not
independent of the operator ${\cal O}_m$, as in (\ref{TequalsO}).
Unfortunately, correlators constructed from the solutions
(\ref{symh}), (\ref{symH}) seem to suffer from the same pathologies as
(\ref{badcorr}).

The reader may justifiably complain that two-point functions
associated to tensor fluctuations should require a more complex
treatment than the inert scalar case, since the actions
are different.  This is further evidence that the procedure
(\ref{InertCorr}) must be generalized in the case of the
graviton/active scalar system.  In section \ref{2point}, we evaluate
the full gravity + scalar action to second order in fluctuations, in
the hope of finding such a generalization.

\section{The Coulomb branch}
\label{coulomb}

Flows involving a single scalar from the ${\bf 20'}$ of $SO(6)$,
which preserve 16 of the 32 bulk supercharges, were
discussed in \cite{fgpw2,bs}.  
Unlike the flow involving $m$, which
is dual to an operator deformation of the Lagrangian of ${\cal N} =4$
SYM, these flows change the vacuum of the field theory, moving it out
onto the Coulomb branch.  The distinction can be perceived by
examining the scaling of the scalar profile near the boundary.  The
${\bf 20'}$ scalars are dual to operators with $\Delta = 2$, and the
Coulomb branch profiles scale as $z^2$ rather than the $z^2 \ln z$
associated with an operator flow.

The five flows discussed in \cite{fgpw2} all have known lifts to 10D
configurations, namely the geometries produced by discs of D3-branes
of various dimensionalities.  There are three inequivalent single
scalars in the ${\bf 20'}$, and two of these have distinct flows in
the positive and negative directions in field space, for five in all.
We consider the flow called $n=2$ in \cite{fgpw2}, and examine
fluctuations of two inert scalars and the graviton/active scalar
system.

\subsection{The background flow}

The scalar field $\varphi$ involved in the $n=2$ Coulomb branch flow
also played a role in a two-scalar flow to an ${\cal N}=1$
superconformal fixed point \cite{fgpw1}, where it was denoted
$\varphi_3$.  The other scalar, there called $\varphi_1$, is a member
of the same ${\bf 10} \oplus {\bf \overline{10}}$ Weyl orbit as
$\sigma$.  The potential and superpotential of this two-scalar
subspace is
\begin{eqnarray}
W(\varphi, \sigma) = - \frac{1}{4} \, e^{-2 \varphi /\sqrt{6}} \left[
e^{\sqrt{6} \varphi} (3- \cosh(2 \sigma)) + 2 (\cosh(2 \sigma) +1)
\right] \,, \\ \label{Coulombpot}
V(\varphi, \sigma) = - \frac{g^2}{4} \, e^{2 \varphi/\sqrt{6}} \, \left[
e^{-\sqrt{6} \varphi} \left( \frac{3}{4} + \frac{1}{2} \cosh(2 \sigma)
- \frac{1}{4} \cosh^2(2 \sigma) \right) + (1+ \cosh(2 \sigma)) + \right. 
\\  \left.
 \frac{1}{16} e^{\sqrt{6} \varphi} ( 1- \cosh^2 (2 \sigma) )
 \right] \,. \nonumber
\end{eqnarray}
In the notation of \cite{fgpw2}, $\varphi = - \mu$ is the active
scalar in the $n=2$ Coulomb branch flow, while $\sigma$ is inert.  The
superpotential for $\varphi$ alone is
\begin{eqnarray}
\label{coulW}
W(\varphi) = - e^{-2 \varphi/\sqrt{6}} - \frac{1}{2} \,
e^{4 \varphi/\sqrt{6}} \,.
\end{eqnarray}
The solution to the equations (\ref{First}) involving (\ref{coulW})
can be obtained,
but they are  not of direct
use since it is more convenient to use a radial coordinate that is a 
function of the scalar itself:
\begin{eqnarray}
v \equiv e^{\sqrt{6} \, \varphi} \,.
\end{eqnarray}
The boundary is at $v=1$.  For this flow $\varphi \rightarrow -
\infty$, so $v \in [0,1]$ with a curvature singularity at $v_s = 0$.
We have the relations
\begin{eqnarray}
\nonumber
W &=& - \frac{1}{2} \,\frac{v + 2}{v^{1/3}} \,, \quad \quad
W' = \frac{2}{\sqrt{6}} \, \frac{1-v}{v^{1/3}} \,, \\
W'' &=& -\frac{2}{3} \, \frac{1+2v}{v^{1/3}} \,, \quad \quad
W''' = \frac{4}{3 \sqrt{6}}  \frac{1 - 4 v}{v^{1/3}}\,, \\
e^{2A} &=& \frac{\ell^2}{L^2} \, \frac{v^{2/3}}{1-v} \,, \quad \quad
\frac{\partial v}{\partial r} = \frac{2}{L} \, v^{2/3} \, (1-v) \,. \nonumber
\end{eqnarray}
One can calculate the horospheric variable $z$ in terms of $v$
and the relationship is
\begin{eqnarray}
\label{horov}
v = \mbox{sech}^2 \left(\frac{z\, \ell}{L^2} \right) \,.
\end{eqnarray}
Following \cite{fgpw2} we have introduced the length $\ell$, the
radius of the disc of D3-branes in ten dimensions.  Taking $\ell/L
\rightarrow 0$ with $z/L$ fixed removes the flow and restores pure
anti-de~Sitter space.  This definition is reminiscent of that for the
variable $u$ in the ${\cal N}=1$ flow (\ref{udef}); one difference is
that for (\ref{horov}) the singularity is at $z \rightarrow \infty$,
and is thus at infinite proper distance.

\subsection{Correlators of inert scalars}

The two-point function of the dilaton $\phi$ has been calculated
previously in this background \cite{fgpw2}.  The Klein-Gordon equation
becomes
\begin{eqnarray}
\label{coulombphieqn}
\phi''(v) + \frac{2-v}{v(1-v)} \phi'(v) + \frac{p^2 L^4}{4 \ell^2} \frac{1}{v^2(1-v)} \phi(v) = 0\,,
\end{eqnarray}
which has the solution 
\begin{eqnarray}
\phi(v) = v^a F(a,a;2+2a;v) \,,
\end{eqnarray}
where
\begin{eqnarray}
\label{a}
a \equiv  -\frac{1}{2} + \frac{1}{2}\sqrt{1 - \frac{L^4p^2}{\ell^2}}  \,.
\end{eqnarray}
This solution is regular at the singularity for spacelike $p^2$; the
second solution of (\ref{coulombphieqn}) has a leading $v^{-a-1}$ and
is not regular. The result for the 2-point function calculated
 from (\ref{2ptfunc}) is

\begin{eqnarray}
\label{coulombdilsoln}
\langle{\cal O}_{\phi}(p) \, {\cal O}_{\phi}(-p)\rangle = - \frac{1}{4} p^4 \psi \left( \frac{1}{2} +
\frac{1}{2}\sqrt{1 - \frac{L^4p^2}{\ell^2}} \right) \,,
\end{eqnarray}
which as remarked in \cite{fgpw2} has a branch cut along the positive
real axis, indicating a continuous spectrum with gap or threshold at 
$m_{gap}^2 =\ell^2/L^4$.

For the $\sigma$ field, there is a $U(\varphi)$ term in the equation
of motion (\ref{inerteq}) owing to the potential (\ref{Coulombpot}),
\begin{eqnarray}
U(\varphi) \equiv \left. \frac{\partial^2 V(\varphi,\sigma)}{\partial
\sigma^2} \right|_{\sigma = 0} = - \frac{g^2}{4} \, e^{2
\varphi/\sqrt{6}}  \left[4 - e^{\sqrt{6} \varphi} \right] \,,
\end{eqnarray}
and the wave equation becomes
\begin{eqnarray}
\sigma''(v) + \frac{2-v}{v(1-v)} \sigma'(v) + \frac{1}{v(1-v)} \left(  \frac{p^2 L^4}{4 \ell^2} \frac{1}{v} + \frac{(4-v)}{4(1-v)} \right) \sigma(v) = 0\,,
\end{eqnarray}
which has the regular solution
\begin{eqnarray}
\sigma(v) = v^a \sqrt{1-v} F(a,a+1;2+2a;v) \,,
\end{eqnarray}
with $a$ as (\ref{a}).  The correlation function is
\begin{eqnarray}
\langle {\cal O}_{\sigma}(p) \, {\cal O}_{\sigma}(-p) \rangle = p^2 \psi \left(
\frac{1}{2} + \frac{1}{2}\sqrt{1 - \frac{L^4p^2}{\ell^2}} \right) \,,
\end{eqnarray}
This correlator also has a continuous spectrum, with the same gap as
(\ref{coulombdilsoln}).  The momentum-dependence is appropriate for an operator
of $\Delta = 3$.

We may check that this behavior is consistent with the SUSY QM
viewpoint.  Calculating the Schr\"odinger potentials ${\cal
V}_{\phi}(z)$ and ${\cal V}_{\sigma}(z)$ (\ref{schpot}), we find that
${\cal V}\rightarrow \infty$ at the boundary for both, while they
asymptote to $\ell^2/L^4$ in the deep interior.  This is precisely the
expected behavior for a potential with a continuous spectrum and a gap
at $m_{gap} = \ell/L^2$.  (For the dilaton, this was already noticed
in \cite{fgpw2}.)

As always, the dilaton wave equation can be rewritten in Schr\"odinger
form with an exact SUSY QM potential given by the prepotential
(\ref{universaldil}).  Normalizability fails near the boundary for
the dilaton zero-mode. As in the ${\cal N}=1$ case, we find that the
Coulomb branch flow is along a ridge of the potential (\ref{Coulombpot})
in the $\sigma$-direction, and
thus $U(\varphi)$ contributes negatively to the $\sigma$ Schr\"odinger
potential (\ref{schpot}).  However, again we find that the potential
can be cast into exact SUSY QM form, this time with the modified
prepotential
\begin{eqnarray}
{\cal U}_{\sigma}(z) = - \frac{1}{2} g \, e^{A} W \,.
\end{eqnarray}
Unlike the ${\cal N}=1$ case (\ref{symsigmasusypot}), this doesn't
have a new term beyond that of (\ref{universaldil}), but the
coefficient is modified instead.  In this case normalizability of
the zero-mode fails at the singularity.

\subsection{The graviton/active scalar system}
\label{coulombactive}

One can show that (\ref{second}) becomes in this case
\begin{eqnarray}
\label{coulombsecond}
R''(v) + \frac{2}{v} R'(v) + \frac{p^2 L^2}{\ell^2} \,
\frac{1}{v^2(1-v)} R(v) = 0 \,.
\end{eqnarray}
This particularly simple form results because the potential-type term
$(2/3) W W'' + (8/9) W^2 + (1/2) W' W''' - W'^2$ evaluates to zero; we
have no ready explanation for this cancellation. The relevant solution
to (\ref{coulombsecond}) is
\begin{eqnarray}
\label{coulombR}
R(v) = v^a \, (1-v) \, F \left( 1+a, 2+a; 2+2a; v \right) \,.
\end{eqnarray}
The second solution also has $v^{-a-1}$ behavior and cannot be made
regular in the interior.  One can integrate (\ref{coulombR}) to obtain the
solution for $\varphit$:
\begin{eqnarray}
\label{3f2}
\varphit(v) = v^a \, (1-v) \; _3F_2 \left( 1+a, 2+a, \frac{1}{3} +a; 2+2a,
\frac{4}{3} +a; v \right) \,.
\end{eqnarray}
We could add a multiple of the universal solution to (\ref{3f2}), but this
would destroy regularity at the interior singularity at $v=0$.

One can see that (\ref{coulombsecond}) has a constant zero-mode solution,
and that (\ref{coulombR}) reduces to a constant as $p^2 \rightarrow 0$ which is
the same as $a \rightarrow 0$. Although this constant mode is not
normalizable,
one might suspect that supersymmetric quantum mechanics is again at work.
Indeed one can again look back to (\ref{second}) in the present case where
the
potential-type term vanishes. The equivalent Schr\"odinger equation is then
supersymmetric with prepotential $U(z) = g/2 \, e^A \, (W''- 5W/3)$.
 
We may use (\ref{algH}),(\ref{algh}) to determine the graviton modes
associated with (\ref{3f2}):
\begin{eqnarray}
\label{coulombh}
h &=& - \frac{4 \sqrt{2} v^a}{3 \sqrt{3} L^4 p^2 (a+1/3)} \left[ 4
\ell^2(1+3a) (-2+4v+v^2 + a(-2+v+v^2)) \; \times \right. \\ 
&& F(1+a,2+a;2+2a;v) - 2 \ell^2 v(1-v)(2+v) (2+7a+3a^2) F(2+a,3+a,3+2a,v)
\nonumber\\   &-&3  \left. (2+v) L^4 p^2 \: _3F_2 \left( 1+a, 2+a,
\frac{1}{3} +a; 2+2a, \frac{4}{3} +a; v \right) \right]  \,, \nonumber \\
\label{coulombH}
H' &=& - \frac{ \sqrt{2} v^{a-1/3}(1-v)}{3 \sqrt{3} L \ell^2 p^2
(a+1/3)} \left[ 4 \ell^2(1+3a) (-2+5v + 3a(v-1)) \; \times \right. \\ 
&& F(1+a,2+a;2+2a;v) +6 \ell^2 v(1-v) (2+7a+3a^2)
F(2+a,3+a,3+2a,v) \nonumber\\ &+& \left. 3 L^4 p^2 \: _3F_2 \left(
1+a, 2+a, \frac{1}{3} +a; 2+2a, \frac{4}{3} +a; v \right) \right] \,.
\nonumber
\end{eqnarray}
Our analysis of the fluctuations (\ref{3f2}), (\ref{coulombh}),
(\ref{coulombH}), all of which contain the generalized hypergeometric
function $_3F_2$, is hampered because the relevant analytic
continuation formulae are not in the literature.  In particular we
need the expansion of (\ref{3f2}) near the boundary, {\em i.e.}\ a
series in $(1-v)$. We proceed by expanding (\ref{coulombR}) in series
first, and then integrate to obtain a series for $\varphit$.  Doing
this we see the leading logarithmic singularity:
\begin{eqnarray}
\label{3f2series}
\varphit = \frac{(a+1/3) \Gamma(2+2a)}{\Gamma(1+a)\Gamma(2+a)}(1-v) \left(
- \ln(1-v) + \varphit_0(a)  +{\cal O}(1-v) \right) \,,
\end{eqnarray}
and while all ${\cal O}(1-v)$ terms could be calculated from higher
order terms in the expansion of (\ref{coulombR}), the integration
constant $\varphit_0$ remains undetermined.  In principle it is fixed
by our choice not to add a multiple of the universal solution to
(\ref{3f2}), but it is nontrivial to calculate it.

Let us consider $h$.  We expect a graviton mode not to be singular on
the boundary, and indeed $1/(1-v)$ and $\ln(1-v)$ terms cancel between
the various terms in (\ref{coulombh}).  The series expansion of $h(v)$
is then
\begin{eqnarray}
\label{coulombhseries}
h(v) = - \frac{16 \sqrt{2} \ell^2}{ \sqrt{3} L^4 p^2} \frac{
\Gamma(2+2a)}{\Gamma(1+a)\Gamma(a)} \left( 3 \varphit_0(a)
-\frac{2}{a(1+a)} + 6 \gamma +6\psi(1+a) \right) + {\cal O}(1-v)^2\,.
\end{eqnarray}
What form do we expect for $h(v)$?  In the field theory, conformal
invariance is only spontaneously broken, and consequently
$T^i_i = 0$ continues to hold as an operator equation 
on the Coulomb branch. As a result, one might expect $h(v)$ to fall off
more rapidly on the boundary than in the case of an operator deformation,
so that it does not excite a dual field theory operator.  
A hint of this behavior is
present in the fact that no ${\cal O}(1-v)$ term appears in
(\ref{coulombhseries}).  We are therefore led to postulate that the
constant term vanishes as well, and
\begin{eqnarray}
\label{intconst}
\varphit_0(a) = \frac{2}{3a(1+a)} -2 \gamma -2\psi(1+a) \,.
\end{eqnarray}
This is a tempting assumption, since (\ref{3f2series}) then produces
the two-point function\footnote{Formulas (\ref{2ptfunc}), (\ref{pureadscorr})
 are modified for a dimension $\Delta=2$
operator. See (25) of \cite{mr}.}
\begin{eqnarray}
\label{goodcorr}
\langle {\cal O}_{\varphi}(p) \, {\cal O}_{\varphi}(-p)\rangle \, =
\psi \left(
\frac{1}{2} + \frac{1}{2}\sqrt{1 - \frac{L^4p^2}{\ell^2}} \right) \
    + \frac{4L^4}{3p^2 \ell^2}
\end{eqnarray}
In addition to the expected branch point there is a pole at $p^2 =
0$. This is a problem for the formalism since the associated constant
zero-mode is non-normalizable, so it should not show up in the
correlator. It is also a problem for the field theory
interpretation. To see this recall that an extra factor of $N^2$ must
be inserted to agree with the short distance form $\langle {\cal
O}_{\varphi}(x) \, {\cal O}_{\varphi}(0)\rangle \;\; \sim \: N^2/x^4$
in field theory. Although the field theory contains massless Goldstone
bosons from the breaking of $SO(6)$ flavor symmetry to $SO(4) \times
SO(2)$ by the disc of D3-branes, it appears very unlikely that these
states should couple to ${\cal O}_{\varphi}= {\Tr} X^2$ with strength
$N$\footnote{We thank S.~Gubser and A.~Hanany for discussions of this
issue.}.

This motivates us to test the assumption (\ref{intconst}) numerically.
Specifically we took the defining power series $_3F_2= \sum_n c_n v^n$
about $v=0$ which is logarithmically divergent at $v=1$ and subtracted
the explicitly summable series of the large $n$ limit of the $c_n$
(obtained using Stirling's formula). The result is a convergent series
whose value at $v=1$ gives the unknown constant ${\varphit_0(a)}$. The
numerical series agrees remarkably well with (\ref{coulombhseries})
for parameter values $a \gg 1$, but agreement fails for small $a$ since
the poles of the power series coefficients for $a \leq 0$ do not
coincide with those of (\ref{coulombhseries}).  We must therefore
conclude that the physically motivated assumption above does not agree
with the properties of $_3F_2$. It should also be observed that
the same correlator (\ref{goodcorr}) can be obtained by subtracting a
multiple of the universal solution from $h(v)$ and $\varphit(v)$, so as
to impose the condition $h(1)=0$.
This diffeomorphism has exactly the effect of inserting
(\ref{intconst}) in (\ref{3f2series}). However the diffeomorphism also
makes $h(v)$ and $\varphit(v)$ singular at the origin, and we cannot
presently justify it.

The situation may be summarized as follows. We imposed the condition
that $h(v)$ vanish on the boundary because it seems to be physically
required for Coulomb branch flows. This leads to a correlation function
with apparently unphysical zero-mode poles. Further, the condition
$h(1)=0$ does not seem to be compatible with the properties of $_3F_2$
obtained by numerical study. Perhaps the analytic form of the constant
$\varphit(a)$ could illuminate the situation.

\section{Calculation of correlation functions}
\label{2point}

Our main purpose here is to discuss some of our attempts to determine
the correct prescription for the calculation of active scalar and
graviton correlators. First we will outline a calculation of the
on-shell supergravity action (\ref{Action}) through second order in
the fields $\varphit$, $h$, and $H$ which are coupled by the
fluctuation equations of Sec \ref{equations}. Because of the coupling,
a single set of boundary data determines the boundary behavior of all
the fields.  Accordingly, factors of $h$ and $H$ in the action will
contribute to the correlation functions of the active scalars. Hence
to calculate two-point functions we should keep terms quadratic in any
of $\varphit$, $h$ and $H$.  One then varies the on-shell action with
respect to the boundary data to obtain the correlation function.

The bulk action is given in (\ref{Action}).  In addition, it is
well-known that this should be supplemented with certain boundary
terms \cite{liutseytlin,frolov}.  First among these is the 
Gibbons-Hawking term, which is
included to generate a well-posed Hamiltonian formalism \cite{gh}:
\begin{eqnarray}
\label{ghterm}
S_{GH} &=& \frac{1}{2} \int_{\partial} d^4x \sqrt{g_4} \, {\cal K} =
\frac{1}{2} \int_{\partial} d^4x \sqrt{g_4} \, \nabla_{\mu} n^{\mu} =
\frac{1}{2} \int_{\partial} \partial_r \sqrt{g_4} \\ &=& \frac{1}{2}
\int_{\partial} e^{4A} \, \left[ \frac{1}{2} h^i_i{}' + \frac{1}{4} h^i_i
h^j_j{}' - \frac{1}{2} h^{ij} h_{ij}' + 4A' \left(1+ \frac{1}{2} h^i_i +
\frac{1}{8} (h^i_i)^2 - \frac{1}{4} h^{ij} h_{ij} \right) + \cdots
\right] \,, \nonumber 
\end{eqnarray}
where ${\cal K} = \nabla_{\mu} n^{\mu}$ is the trace of the second
fundamental form of the boundary, with $n^{\mu}$ a normal to the
boundary; in the second line we expand this term to second order in
the fluctuations.  In addition, a boundary cosmological term has been
recommended for canceling the leading volume divergence:
\begin{eqnarray}
\label{volume}
S_{vol} &=& \beta \int_{\partial} \sqrt{g_4} \, A' \\ &=& \beta
\int_{\partial} e^{4A} \, A' \, \left( 1 + \frac{1}{2} h^i_i +
\frac{1}{8} (h^i_i)^2 - \frac{1}{4} h^{ij} h_{ij} + \cdots \right) \,,
\end{eqnarray}
for some $\beta$ which we will fix momentarily.  Note that $A' = 1/L$
is a constant on boundary, which we include explicitly to simplify the
formulas that follow.

We proceed to expand the bulk action to second order in fluctuations,
making use of the equations of motion as necessary.  After a
remarkably tedious calculation, we find that as in simpler cases, the
bulk action can be reduced entirely to a set of boundary terms:
\begin{eqnarray}
S_{bulk} &=& S_0 + S_1 + S_2 \,, \nonumber \\
S_0 &=& \int d^4x \, e^{4A} \left (- \frac{1}{2} A' \right)\,, \\
S_1 &=& \int d^4x \, e^{4A} \left(- \frac{1}{4} h^i_i{}' - \frac{1}{4} A' h^i_i - \varphi' \varphit \right) \,, \nonumber \\
S_2 &=&
\int d^4x \, e^{4A}\left( - \frac{1}{2} \varphit \varphit' - \frac{1}{4}
\varphi' \varphit h^i_i + \frac{3}{16} h^{ij} h_{ij}' - \frac{1}{16}
h^i_i (h^j_j)' + \frac{1}{8} A' h^{ij} h_{ij} - \frac{1}{16}
A' h^i_i h^j_j \right) \,, \nonumber
\end{eqnarray}
evaluated on the boundary $r=R$.
Here we have organized the action by the order of the fluctuations.
The zeroth-order term is the volume divergence, which also receives
contributions from the boundary terms (\ref{ghterm}) and
(\ref{volume}); it is canceled by the choice $\beta = 1/2$, which in
fact removes all terms proportional to $A'$.  The total action then
reduces to 
\begin{eqnarray}
S_{tot} = \int d^4x \, e^{4A} \, \left(- \varphi' \varphit -
\frac{1}{2} \varphit \varphit' - \frac{1}{4} \phi' \varphit h^i_i +
\frac{1}{16} h^i_i h^j_j{}' - \frac{1}{16} h^{ij} h_{ij}' \right) \,.
\end{eqnarray}
We can express this in momentum space in terms of $h$ and $H$,
\begin{eqnarray}
\label{totalaction}
S_{tot} &=& - \varphi'(R) \, \varphit(R,p=0) + \int d^4p \, e^{4A} \,
 \left( - \frac{1}{2} \varphit(R,p) \, \varphit'(R,p)\right. \\ &+&
 \left. \frac{3}{32} \, h(R,p) \, h'(R,p) +\frac{3}{32} \, p^2 H(R,p)
 \, h'(R,p) + \frac{3}{64} \, p^2 h(R,p) \, H'(R,p) \right) \,.
 \nonumber
\end{eqnarray}
The first term in (\ref{totalaction}) is linear in the fluctuation
$\varphit$, and is thus suggestive of a one-point function. It does
not clearly discriminate between the ${\cal N}=1$ flow, where no
one-point function is expected, and the Coulomb branch flow, where
one is. We find this puzzling.

On the other hand, the quadratic terms suggest a modified calculation
of the scalar correlator in which $h$ and $H$ are related to the
boundary data for $\varphit$.  This is straightforward but
complicated.  We have performed such a calculation for the case of the
${\cal N}=1$ active scalar, but ultimately encountered the same
difficulties as in section \ref{sym}.  The resolution of the problem
presumably involves understanding (\ref{totalaction}) better, but
other pieces of the puzzle may still be missing.

A further uncertainty is the issue of the diffeomorphism
invariance.  We refer particularly to (\ref{diffeo}) which has been
interpreted as describing \cite{tanaka,gkr} bending of the
cutoff surface and horizon. It is unsettling that the calculation of
the active correlator in the Coulomb branch flow was so markedly
changed by such a diffeomorphism.  The on-shell action must be
diffeomorphism invariant, and it is not clear to us that this is
manifest in (\ref{totalaction}).  In particular, a term $H$ with no
derivatives appears; this quantity is absent from the equation of
motion and its constant term is thus not determined, but has the form
of a pure diffeomorphism (\ref{4ddiffeo}).  A better understanding
will have to involve coming to terms with these issues.

\section*{Acknowledgements}

We are grateful for discussions with D.~Anselmi, S.~Gubser, A.~Hanany,
A.~Karch, J.~Minahan, G.~Moore, J.~Polchinski, L.~Rastelli, and
N.~Warner.  The research of O.D.\ was supported by the U.S.\
Department of Energy under contract \#DE-FC02-94ER40818.  The research
of D.Z.F.\ was supported in part by the NSF under grant number
PHY-97-22072.


\end{document}